# CRYSTAL CHANNELLING IN ACCELERATORS*

V.M. Biryukov[#], Institute for High Energy Physics, Protvino, Russia

*Abstract*

Crystal lattice can trap and channel particle beams along major crystallographic directions. In a bent crystal, the channelled particles follow the bend. This makes a basis for an elegant technique of beam steering by means of bent channelling crystals, experimentally demonstrated from 3 MeV to 1 TeV. This technique was strongly developed in recent studies at CERN, FNAL, IHEP, and BNL, and can lead to interesting applications also at the LHC, such as crystal collimation making a collider cleaner by an order of magnitude. We review recent developments in the field and show outlook for the future.

## INTRODUCTION

The idea to deflect proton beams using bent crystals, originally proposed by E.N. Tsyganov in 1976 [1] and demonstrated first in Dubna [2] on protons of a few GeV, has received strong development since then [3-6].

Leaving aside the details of channeling physics, accelerator physicist will find many familiar things here:

- Channeled particle oscillates in a transverse nonlinear field of a crystal channel, which is similar to the *betatron oscillations* in accelerator, but on a much different scale (3 μm wavelength at 1 GeV in Si ).
- The crystal nuclei arranged in crystallographic planes represent the "*vacuum chamber walls*". Notice the "vacuum chamber" size of ~2 Angstroms.
- The well-channeled particles are confined far from nuclei (from "aperture"). They are lost then only due to scattering on electrons. This is analogy to "*scattering on residual gas*".
- Like the real accelerator lattice may suffer from *errors of alignment*, the real crystal lattice may have dislocations too, causing particle loss.
- Accelerators tend to use superconducting magnets. Interestingly, the crystals cooled to *cryogenic temperatures* are more efficient, too [7].

## THE EXPERIMENTS AT CERN SPS

Important milestone in 1991 was a bending experiment on H8 beam at CERN where crystal bent first 10% [8] and later up to 60% [6,9] of all beam incident at crystal - orders of magnitude higher than any previous figure (<<1% typically), Fig. 1. The experiment continued [6] with many crystal types and particle species: protons, pions, and ions of high energy, and lead to application for beam splitting in the K12 beam for NA48 at CERN [10].

Next major milestone was a crystal experiment on proton extraction from the SPS [11-13]. For the first time, extraction efficiency was measured as 10%, again orders of magnitude higher than any previous figures.

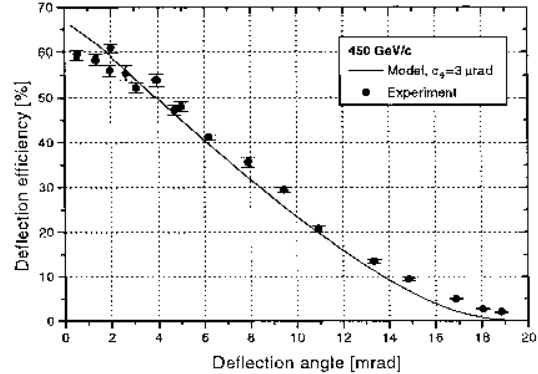

Figure 1. Efficiency vs bending angle for Ge(110) crystal.

The SPS studies helped a lot in understanding crystal extraction. In the framework of SPS studies, computer simulations from the first principles gave an adequate prediction for the experiments. Simulations [14-16] of extraction process included multiple encounters with the crystal, and turns in the accelerator, of the beam particles. Tracking of particle through a bent crystal lattice requires not only a calculation of a particle dynamics in its nonlinear field, but also a generation of random events of scattering on the crystal electrons and nuclei [17].

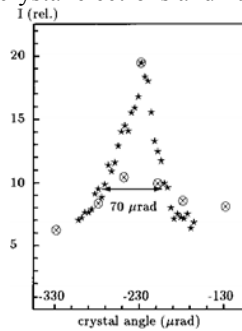

Figure 2. The angular scan of extraction with a U-shaped crystal [18]. Predicted (1993) and measured (1994).

Simulation results were found consistent with the observations in the assumption of a few-μm surface irregularities (`septum width') that suppress channelling on the first pass in crystal. Another SPS experiment used a crystal with an amorphous layer at the edge to suppress channeling in the first passage [12]. The extraction efficiency with this crystal was indeed of the same order of magnitude as found without an amorphous layer, thus confirming the expectation [14-15]. Further extraction studies [12,13] at the SPS continued with crystals of new geometry (``U-shaped"). Fig. 2 [18] shows the first measured [19] angular scan (70 μrad FWHM) of crystal extraction, in agreement with prediction [14].

The dependence of crystal extraction on proton energy was measured at the SPS with the same crystal [13], Fig. 3. This behaviour was well understood in simulations

*Supported by RFBR 06-02-26791 and INTAS-CERN 03-52-6155
[#] http://mail.ihep.ru/~biryukov/

[16,20]. This function, and even the absolute figures of efficiency of multi-pass extraction, can be understood in the framework of analytical theory of crystal extraction [21]; its predictions are shown in Fig. 3.

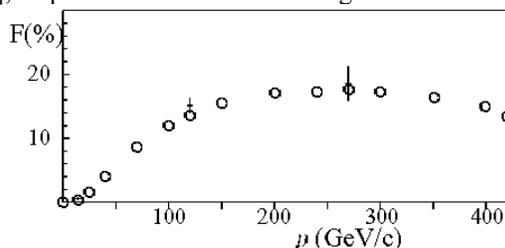

Figure 3 Extraction efficiency vs proton momentum. Analytical theory (o) and SPS data (crosses).

More bending and extraction experiments at CERN SPS were done with Pb ions of highest energy, up to 400 GeV/u [6,22]. These studies have shown that crystal technique is fully applicable to heavy ions. Extraction efficiency observed at CERN SPS with ions was 4-11%.

For the lifetime of crystal, CERN experiment [6] with 450 GeV protons showed that at the achieved irradiation of $5 \cdot 10^{20}$ proton/cm$^2$ the crystal lost only 30% of its deflection efficiency, which means about 100 years lifetime in the intense beam of NA48 experiment.

## THE TEVATRON EXPERIMENT

The Tevatron E853 extraction experiment has provided another check of the technique at a much higher energy of 900 GeV. During the FNAL test, the halo created by beam-beam interaction in the periphery of the circulating beam was extracted from the beam pipe without measurable effect on the background seen by the experimental detectors. The crystal was channeling a 900-GeV proton beam with an efficiency of 25-30% [23,24], showing a rather good agreement with the theoretical expectation. Simulation predicted the efficiency of 35% for a realistic crystal in the Tevatron experiment [25].

Apart from observing the channeled particles, this experiment has measured also the particles dechanneled from the crystal, appearing as a tail. The number of particles in the visible tail was measured 20% of the peak [23]. A simulation of the experiment predicted 25% [25].

## IDEAS FOR EFFICIENT CHANNELING

The crystal length used in the SPS and Tevatron was optimal to bend protons with a *single* pass. The efficiency of the *multi*-pass extraction is defined by the processes of channeling, scattering, and nuclear interaction in the crystal, which depend essentially on the crystal length *L*.

In order to let the circulating particles encounter the crystal many times and suffer less loss in the crystal, one can minimise crystal length to a limit set by channeling physics in a strongly bent crystal [26,27]. This optimisation was studied in simulations in general and for the experiments at SPS and Tevatron [15]. Fig. 4 shows that efficiency in SPS at 120 GeV more than doubles [14] at a new optimum, L≈0.7 cm, w.r.t. a 4 cm crystal. Predicting this boost in efficiency was not trivial: Fig. 4 shows also the absolute efficiency from another simulation [16], predicting just 15% rise The same Figure shows the SPS data with a 4-cm crystal [12].

Similarly, for the Tevatron E853 experiment it was found that extraction efficiency could be increased to ~70% with a crystal length cut to 0.4-1 cm [15,25].

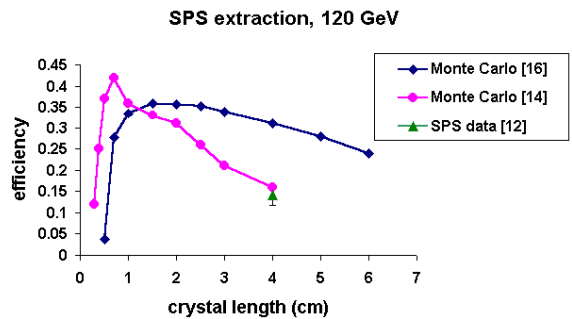

Figure 4. The SPS extraction efficiency vs crystal length. Simulations (1998, 1993) and measurement (1995).

## THE IHEP EXPERIMENTS

Since 1989, IHEP Protvino pioneered a wide practical use of bent crystals as optical elements in high-energy beams for beam extraction and deflection on permanent basis [28]. In 1997, a new extraction experiment was proposed [29] aiming to benefit from very short crystals. Monte Carlo study predicted that crystal can be cut down to ~1 mm along the 70-GeV beam in the extraction set-up of IHEP. This promised tremendous benefits: crystal extraction efficiency could be over 90%. Fig. 5 shows both the predicted [30] extraction efficiency as a function of the crystal length, and some history of IHEP measurements since 1997 [31,32].

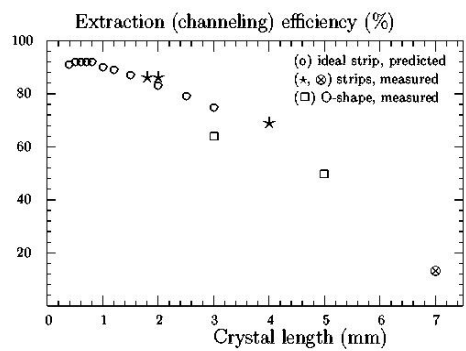

Figure 5. Crystal extraction efficiency for 70 GeV protons. IHEP measurements and Monte Carlo prediction.

Compared to the SPS and Tevatron experiments, the efficiency is improved by a factor of 3-8 while the crystal size along the beam cut by a factor of 15-20 (from 30-40 to 2 mm). It took years in IHEP, and the decisive step was invention of strip-type deflectors [32], very short–down to ~2 mm along the 70-GeV beam. This lead to dramatic boost in crystal efficiency. Now crystal systems extract 70 GeV protons from IHEP main ring with efficiency of 85% (defined as the ratio of the extracted beam to all the beam loss in the ring) at intensity of $10^{12}$. Today, six locations on the IHEP 70-GeV main ring are equipped by crystal extraction systems, serving mostly for routine applications rather than for research. The record efficiency of 85% is pertained even when the *entire beam* stored in the ring is dumped onto the crystal.

Two locations on IHEP ring are dedicated for crystal collimation. There, a bent crystal is upstream of a secondary collimator. Fig. 6 shows the radial distribution of protons measured on the collimator face. It includes the peak of channeled particles bent into the depth of the collimator, and the scattered particles near the edge.

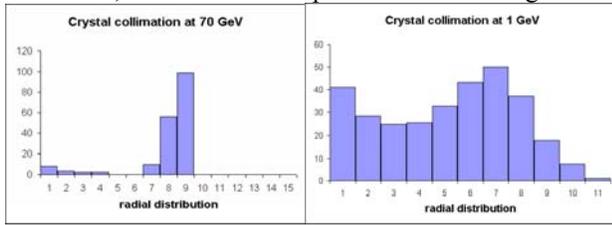

Figure 6. The radial beam profile at the collimator.

At 1.3 GeV with the same crystal, crystal collimation is still quite strong, Fig. 6, although the energy is lowered by two orders of magnitude. The same set-up was tested in a broad energy range in the main ring of U-70. Fig. 7 shows the ratio of the channeled particles to the entire beam dump (the *crystal collimation efficiency*) as measured and as predicted.

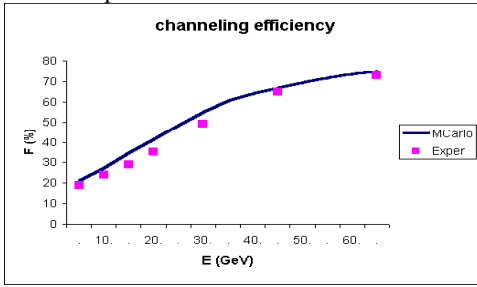

Figure 7. Crystal collimation, ramping energy in U-70.

The background measured downstream of the collimator drops by factor of 2 when the crystal is aligned, Fig. 8. The experiment was world first demonstration of crystal collimation.

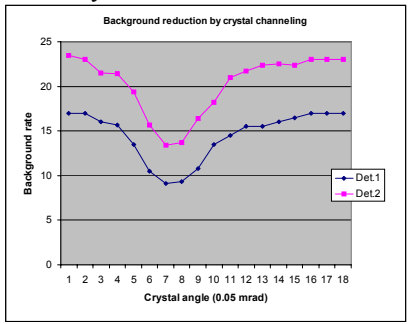

Figure 8. Background versus crystal alignment.

Thermal shock is an issue important for application. In typical IHEP tests, crystal channeled $\sim 10^{12}$ protons (up to $4 \cdot 10^{12}$ in some runs) in a spill of 0.5-1 s duration. Let us illustrate it in the following way. Suppose, all the LHC store of $3 \cdot 10^{14}$ protons is dumped on our single crystal in a matter of 0.2 hour [33]. This makes a beam of $4 \cdot 10^{11}$ proton/s incident on the crystal face. In IHEP, this is just routine work for a crystal, practiced every day. IHEP experience can help broad crystal application to beam optics even at high-intensity machines, e.g. J-PARC [34].

## RHIC CRYSTAL EXPERIMENT

Another experiment on crystal collimation was done at the Relativistic Heavy Ion Collider [35-36]. The yellow ring of the RHIC had a bent crystal collimator of the same type as used in earlier IHEP experiments [31], 5 mm along the beam. By aligning the crystal to the beam halo, particles entering the crystal were bent away from the beam and intercepted downstream in a copper scraper. Beam losses were recorded by the PIN diodes, hodoscope, and beam loss monitors. Fig. 9 shows a typical angular scan from the 2003 RHIC run with gold ions, and predicted angular scan. The two angular curves, measured and predicted, are in reasonable agreement.

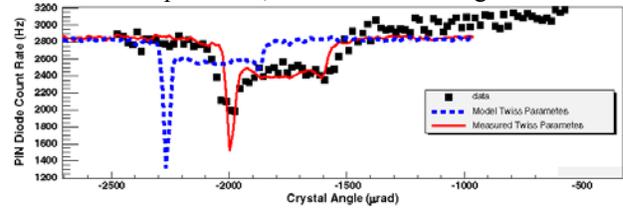

Figure 9 RHIC: nuclear interaction rate as a function of crystal orientation, measured (dots) and simulated with preliminary (blue) and measured (red) optics.

The efficiency is defined as maximum depth of the large dip divided by the background rate. For the 2003 RHIC run, the theory predicted the efficiency of 32%, and the averaged measured efficiency for this run is 26%. The modest figure of efficiency ~30%, both in theory and experiment, is attributed to the high angular spread of the beam that hits the crystal face as set by machine optics. It is worth to compare the efficiency for Au ions at RHIC to the 40% efficiency achieved with similar crystal for protons at IHEP in 1998 [31]. Extraction efficiency observed at CERN SPS with Pb ions was 4 to 11% with a long (40 mm) crystal [22]. The RHIC study was demonstration of world first crystal collimation for heavy ions, with efficiency record high for ions.

## TEVATRON SIMULATIONS

A possibility to improve the Tevatron beam halo scraping using a bent channelling crystal instead of a thin scattering target as a primary collimator was studied at FNAL [37] with realistic simulations using the CATCH, STRUCT and MARS Monte Carlo codes.

It was shown that the scraping efficiency can be increased by an order of magnitude. As a result, the beam-related backgrounds in the CDF and D0 collider detectors can be reduced by a factor of 7 to 14. Calculated results on the system performance taking into account the thickness of near-surface amorphous layer of the crystal are presented in Table 1. Two cases have been compared:
1. The Tevatron RUN-II collimation system with all secondary collimators in design positions, but only one (D17h) horizontal primary collimator in working position. This primary collimator intercepts large amplitude protons and protons with positive momentum deviations.
2. The same collimation scheme, but silicon bent crystal is used instead of primary collimator.

Table 1 The Tevatron: Halo hit rates at the D0 and CDF Roman pots and nuclear interaction rates $N$ (in $10^4$ p/s) in the primary scraper (target or crystal). Simulation [37].

|   | with target | with crystal | | |
|---|---|---|---|---|
|   |   | amorphous layer thickness | | |
|   |   | 10 μm | 5 μm | 2 μm |
| DØ | 11.5 | 1.35 | 1.60 | 1.15 |
| CDF | 43.6 | 5.40 | 3.20 | 3.43 |
| $N$ | 270 | 82.4 | 70.6 | 50.3 |

## SIMULATIONS FOR THE LHC

We evaluated the potential effect of crystal collimation in the LHC using the same computer model [38] already validated with the IHEP, CERN, FNAL, and BNL experiments on channeling. Simulations were done in the LHC both at the collision energy of 7 TeV and at the injection energy of 450 GeV for a nominal beam emittance of 3.75 μm (at 1 σ, defined as r.m.s.). In the model, a bent crystal was positioned as a primary element at a horizontal coordinate of 6σ in the LHC halo, on one of the locations presently chosen for amorphous primary elements of the LHC collimation design [39].

We varied crystal parameters such as the size, bending, alignment angle, material, and the quality of the surface. We observed the efficiency of channeling, i.e., the number of the particles deflected at the full bending angle of the crystal, taking into account many turns in the LHC ring and multiple encounters with the crystal. On the first encounter, the halo particles entered the crystal face within ≤1 μm from the edge. The first 1-μm thick near-surface layer of the crystal was assumed amorphous.

Fig. 10 shows the computed channeling efficiency as a function of the crystal length along the LHC beam for two cases: at flattop (7 TeV) and at injection (450 GeV). The optimal size of the silicon crystal is about 5 mm for 0.1 mrad, and just 3 mm for 0.05 mrad. High efficiency of channeling can be obtained with the same (optimized) crystal both at 7 TeV and at 450 GeV. The efficiency is expected to be 90-94% in the case of crystal bending angle of 0.05-0.1 mrad.

Different bending angles were examined (finding every time the optimal size for the crystal) and the channeling efficiency computed, Fig. 10. If all channeled particles were fully intercepted by the secondary collimator, then only non-channeled particles should contribute to the background in the accelerator. Fig. 10 data means that the halo intensity can be reduced by a factor of 10-25. All the range of crystal deflector size assumed in Fig. 10 is already realised and tested by IHEP in 70 GeV beam.

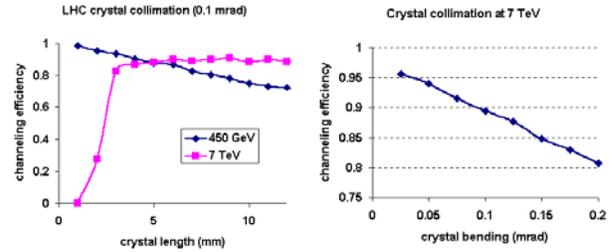

Figure 10. Channeling efficiency in LHC vs crystal length (left) and bending (right).

The optics of traditional collimation and technical considerations may require primary scrapers of different material. The technique of bent crystal channeling is developed also with other materials, e.g. Ge (Z=32) [6] and diamond (Z=6). In simulations [39], comparable efficiencies were obtained with Ge, Si, and diamond at 7 TeV. All these crystals, from diamond to germanium, can serve as an LHC primary scraper. Another interesting (but futuristic) possibility might be the use of nanostructured material [40].

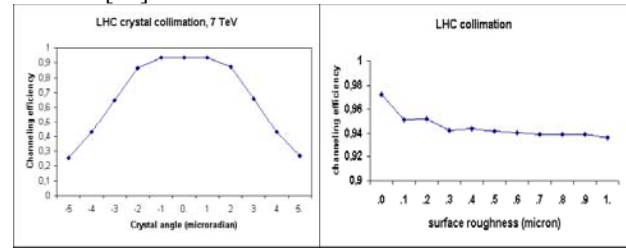

Figure 11. Channeling efficiency vs crystal orientation angle (left) and crystal surface roughness (right).

For efficient operation, crystal must be oriented parallel to LHC beam envelope within ~1 μrad, Fig. 11. We studied also the effect of crystal surface in simulations, Fig. 11. With surface level better than 0.1 μm, the computed efficiency exceeds 97%.

Apart from collimation, more interesting crystal applications are proposed for the LHC, e.g. for CMS and ATLAS calibration by the primary beam of LHC [41].

## STUDIES OF CRYSTAL COLLIMATION AT THE TEVATRON AND CERN SPS

High expectations for crystal collimation at TeV colliders stimulated new experiments at the Tevatron and SPS. The O-shaped crystal tested in RHIC was installed in the Tevatron and tried at 980 GeV in a collimation experiment. Fig. 12 shows the crystal nuclear interaction rate measured (dots) and simulated as a function of crystal orientation [42]. The plot shows a dramatic dip due to channeling with very high efficiency. The measurements and simulations are in good agreement.

A striking feature of the plot is a plateau with the width equal to the crystal bending angle, 0.44 mrad, where the interaction rate is about 50% of that at random orientation. Simulation [43] identified the plateau as a strong effect of beam coherent scattering ("reflection") in a field of bent crystal. The effect of beam "volume reflection" in bent crystals, Fig. 13 (a), was predicted in 1987 [44]. Crystal collimation [35,42,43] gave the first observation of this new physical phenomenon in experiment.

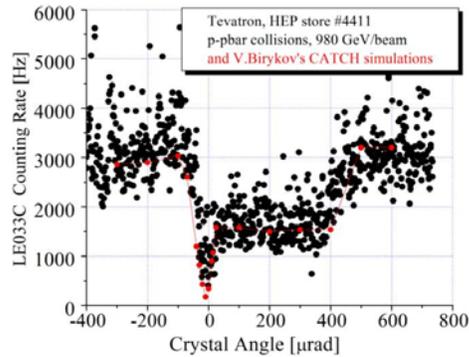

Figure 12. Crystal collimation at 1 TeV.

A new deflector from IHEP, strip-type of 3 mm length and 0.15 mrad bend, is now installed into Tevatron for a collimation test. Simulation predicts a plateau 0.15 mrad wide, with rate suppressed by 65% w.r.t. the random orientation. At the dip, the new crystal should produce a background lower than the O-crystal does, by a factor of 1.5-2 because of channeling with higher efficiency.

Tevatron tests of crystal collimation will provide the best opportunity for validation of the technique; however, lots of new information can be gained from the tests in the SPS. A new experiment at CERN SPS [45] aims to measure directly for the first time a reflection angle of 400 GeV protons in a bent crystal. Fig. 13(b) shows the predicted distribution downstream of the crystal in the SPS beamline: 96% of the beam is reflected with (most probable) angle of 13 μrad.

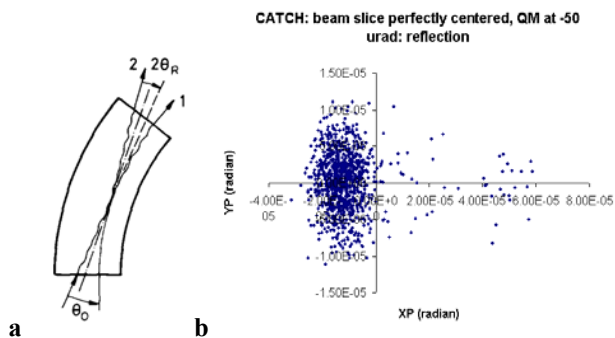

Figure 13 (a) Sketch [22] of a channeled (1) and reflected (2) particle. (b) Predicted volume reflection for the SPS.

## CONCLUSION

Crystal works efficiently, up to 85%, at very high intensities (~$10^{12}$), with a lifetime of many years. Monte Carlo model successfully predicts crystal work in circulating beam, as demonstrated in experiments at up to 1 TeV. The same crystal works efficiently over full energy range, from injection through ramping up to top energy, as demonstrated at IHEP from 1 through 70 GeV and as seen in simulations for the LHC.

Crystal would be very efficient in the LHC environment. The expected efficiency figure, ~90%, is already experimentally demonstrated at IHEP and being confirmed in the Tevatron studies. This would make the LHC cleaner by an order of magnitude.